\documentclass[11pt]{article}  
\usepackage{menuproc}
\usepackage{cite}
\usepackage{epsfig}
\usepackage{amsmath,amssymb}
\def\tens{\otimes}
\def\Id{{\rm 1\kern-.3em I}}

\begin{document}
%
%
%
\titlematter{A relativistic quark model of baryons}%
{Bernard Metsch, Ulrich L\"oring}%
{Institut f\"ur Theoretische Kernphysik,\\
 Universit\"at Bonn, Nu{\ss}allee 14-16, D 53115 Bonn, Germany\\
 \sf E-mail: metsch@itkp.uni-bonn.de, loering@itkp.uni-bonn.de}
{
On the basis of the three-particle Bethe-Salpeter equation we
formulated a relativistic quark model for baryons. Assuming the
propagators to be given by their free form with constituent quark
masses and the interaction kernel by an instantaneous potential, which
contains a string-like parameterization of confinement and a flavor
dependent interaction motivated by instanton effects we can account
for the major features in the baryon spectrum, such as the low
position of the Roper resonance and the occurrence of approximate
parity doublets apparent in the $N$- and $\Lambda$-spectra.
}
%
%

\section{Introduction}

Still the most successful description of the baryonic excitation
spectrum is the non-relativistic constituent quark model (NRCQM). It
essentially assumes that baryon resonances are $q^3$-bound states,
coupled weakly to meson fields. This certainly can be questioned: The
decay widths of hadronic resonances are in general appreciable. The
assumption should be understood to state 
that the coupling to strong decay channels does not
influence the \textit{relative} positions of resonances at the level
of, say, 50 MeV. Thus the major mass splittings are supposed to
reflect the excitation of the degrees of freedom of three constituent
quarks, which have an effective constituent quark mass, carry flavor
and color in the fundamental representations of the corresponding
groups, are described by Pauli-spinors and thus obey the
Pauli-principle. In its simplest form, assuming the interaction to be
given by harmonic forces, these ingredients provides a natural
explanation of the ground state flavor octet and decuplet, the
lowest orbital excitations (``1-$\hbar\omega$-states'') and the rough
position of the lowest state of each spin. A comparison with the
experimental spectrum, however, also reveals some conspicuous
shortcomings: there is no natural explanation of the position of the
lowest scalar/isoscalar excitations, such as the Roper-resonance, nor
of the low position of some radial/orbital $\Delta$-resonances at 1.9
GeV. In addition to these phenomenological considerations there are
also some more fundamental objections: even with constituent quark
masses of a few hundred MeV, quarks in hadrons are not really slow---
the expectation value of $\frac{p}{m_q}$ being roughly of order
unity--- processes at larger momentum transfer intrinsically require a
relativistically covariant treatment of boosts, and in particular for
the ground state mesons there are binding effects which are large
compared to the masses of the constituents.

These considerations led us to the formulation of a relativistically
covariant quark model for mesons on the basis of the instantaneous
Bethe-Salpeter equation, see e.g. \cite{kol00,ric00} and
references therein, while at the same time retaining all the successful
features of the NRCQM.   

\section{The covariant $q^3$ Bethe Salpeter equation}

In momentum space the Bethe-Salpeter equation for the
Bethe-Salpeter amplitude $\chi_{\bar P}(x_1,x_2,x_3) = \langle 0 \mid
T\Psi(x_1)\Psi(x_2)\Psi(x_3) \mid \bar P \rangle$ of a bound state
with $\bar P^2=M^2$ reads \cite{loering01a}
\begin{eqnarray}
\label{eq:1}
\lefteqn{\chi_{\bar P}(p_\xi,p_\eta)= S^F_{1}(\frac{1}{3}\bar
P+p_{\xi}+\frac{1}{2}p_{\eta})\otimes S^F_{2}(\frac{1}{3}\bar
P-p_{\xi}+\frac{1}{2}p_{\eta})\otimes S^F_{3}(\frac{1}{3}\bar
P-p_{\eta})}\nonumber\\ & &\qquad\qquad\qquad\qquad\qquad\times (-i)\;
\int\frac{d^4\!p_\xi'}{(2\pi)^4}\;\frac{d^4\!p_\eta'}{(2\pi)^4}\;
K(\bar P,p_\xi,p_\eta,p_\xi',p_\eta')\; \chi_{\bar P}(p_\xi',p_\eta')\,,
\end{eqnarray}
where $\bar P=\sum_{i=1}^3p_i$ is the total 4-momentum and
$p_\xi=\frac{1}{2}(p_1-p_2)$, $p_\eta=\frac{1}{3}(p_1+p_2-2p_3)$ are
Jacobi momenta. $S^F$ denotes the full quark propagator and  $K =
K^{(3)}+\bar K^{(2)}$ is the
integral kernel which
contains both irreducible three-body $K^{(3)}$ 
and two-body forces $\bar K^{(2)}$.  Both $S^F$ and $K$ are
unknown functions for strongly interacting particles and thus have to
be modeled. We start by making the assumption, that the full quark
propagator can be approximated by its free form
\begin{displaymath}
S^F_i(p)\approx i\left[\gamma(p)-m_i\;+\; i\epsilon\right]^{-1}\,,
\end{displaymath}
where $m_i$ is the effective constituent quark mass, which thus is a
free parameter in our model. We furthermore assume, that the
irreducible kernel can be suitably written as
\begin{equation}
  \label{eq:2}
K^{(3)}(p_\xi,p_\eta;p_\xi',p_\eta')\mid_{\bar P=(M,\vec 0)}\!\!
=
\!\!  V^{(3)}(
p_{\xi\perp},p_{\eta\perp};p_{\xi\perp}',p_{\eta\perp}')\,,
\qquad
K^{(2)}(p_{\xi},p_{\xi}')\mid_{\bar P=(M,\vec 0)}\!\!  
=
\!\!
V^{(2)}(p_{\xi\perp},p_{\xi\perp}')
\end{equation}
where $p_{\perp} \equiv p - \frac{\bar P p}{{\bar P}^2}\;\bar
P\;$. In a frame where $\bar P = (M,\vec 0)$ the kernel
then depends on the spatial components of the relative momenta only and
thus reflects an instantaneous potential in the rest frame of the
baryon. Both assumptions are in fact borrowed from the NRCQM, which
accounts for confinement by assuming a string-like three body
potential for constituent quarks. In fact the instantaneous approach 
is somewhat problematic for two particle interactions. 
For the moment we will assume 
that retardation effects of the propagation of the spectator quark can
be parametrized by construction of an effective 3-body kernel through
$\langle G_0 \rangle V^{\rm eff}_{ij} \langle G_0 \rangle
 := \langle G_0 K^{(2)}_{ij} G_0 \rangle $, where $G_0 :=
S^F_{1}\otimes S^F_{2}\otimes S^F_{3}$ is the free 3-quark propagator 
and $\langle G_0 \rangle:= \int
\frac{dp_\xi^0}{2\pi} \frac{dp_\eta^0}{2\pi} G_0$. With these
assumptions one can integrate out the $p_\xi^0, p_\eta^0$ dependence
in Eq.(\ref{eq:1}) and derive the Salpeter equation (see Ref. \cite{loering01a}):
\begin{equation}
  \label{eq:3}
\Phi = -i \langle G_0 \rangle \;\;( V^{(3)} + {V_{\textrm{eff}}} ) \;\;{\Phi}
\end{equation}
for the Salpeter amplitude $\Phi := \int \frac{dp_\xi^0}{2\pi}
\frac{dp_\eta^0}{2\pi}\,\chi$, with $V_{\textrm{eff}}=\sum_{i<j}
V^{\textrm{eff}}_{ij}$\,.
The normalization of $\Phi$ follows
from the normalization of the Bethe-Salpeter amplitude and reads
\begin{equation}
  \label{eq:4}
\sum_{\alpha\beta\gamma} \int\frac{d^3\!p_\xi d^3\!p_\eta }{(2\pi)^6}
{\Phi}^*_{M\alpha\beta\gamma}(\vec p_\xi,\vec p_\eta)\;
{\Phi}_{M\alpha\beta\gamma}(\vec p_\xi,\vec p_\eta)=2M\,.
\end{equation}
This allows to reformulate Eq.(\ref{eq:3}) as an eigenvalue problem
${\mathcal H}\,\Phi = M\,\Phi$ with the Salpeter Hamiltonian
\begin{eqnarray}
  \label{eq:5}
\lefteqn{(\mathcal{{H}}{\Phi}_M)(\vec p_\xi,\vec p_\eta) $=$ \sum_{i=1}^3
H_i(\vec p_i){\Phi_M}(\vec p_\xi,\vec p_\eta)
%
+ \bigl[
{\Lambda_1^+}(\vec p_1)\otimes{\Lambda_2^+}(\vec
p_2)\otimes{\Lambda_3^+}(\vec p_3)}  
\\ && +
{\Lambda_1^-}(\vec
p_1)\otimes{\Lambda_2^-}(\vec p_2)\otimes{\Lambda_3^-}(\vec
p_3)\bigr]\nonumber
\;\gamma^0\!\otimes\!\gamma^0\!\otimes\!\gamma^0\;
\int\frac{d^3\!p_\xi'}{(2\pi)^3}\;\frac{d^3\!p_\eta'}{(2\pi)^3}\;
(V^{(3)}\!+\!{V_{\textrm{eff}}})(\vec p_\xi,\vec p_\eta,\vec
p_\xi',\vec p_\eta')\; {\Phi}_M(\vec p_\xi',\vec p_\eta')\nonumber
\end{eqnarray}
with the Dirac Hamiltonian $H_i$ and $\Lambda_i^{\pm}$ being projectors
on states of positive ($+$) and negative ($-$) energy spinors. 
Note that with the
present instantaneous approximation we get the same number
 of states
as in the NRCQM, nevertheless our approach goes beyond a simple
''relativization'' by explicitly taking into account the coupling to
negative energy components. In this formulation we can
determine masses and Salpeter amplitudes for baryons by diagonalizing
the Hamiltonian of Eq.(\ref{eq:5}) in an appropriate large, but finite
basis. In order to calculate electroweak observables in the framework
of the Mandelstam formalism we need the full
Bethe-Salpeter amplitude. This can be found in the rest frame through
the Bethe Salpeter equation (see Ref. \cite{kretzsch01})
\begin{displaymath}
\chi_M = -i G_0 (V^{(3)}_{\mbox{\footnotesize conf}} 
+ V^{\mbox{\footnotesize eff}} ) {\Phi}\,,
\end{displaymath}
and, because of formal covariance, can be calculated 
for any on-shell momentum $\bar P$ with $\bar P^2= M^2$\,.
\section{A covariant quark model for light-flavored baryons}
In order to calculate the spectrum of baryons with light flavors we
have to specify the interaction kernels $K^{(3)}=V^{(3)}$ and
$K^{(2)}=V^{(2)}$ in instantaneous approximation.
We parameterize a three-body confinement kernel by a potential
$V^{(3)}_{\rm conf}$ which rises linearly with inter-quark distances.
The gross features of the baryon resonances seem to indicate that the
dominating confinement forces should be spin-independent, at least in
the non-relativistic limit; moreover, too large spin-orbit effects
should be suppressed.  To realize both properties we assume the
Dirac structure given by a specific combination of spin-independent
scalar and time-like vector\footnote{The appearance of
$\gamma^0\tens\gamma^0$ is allowed because the potential is defined in
the rest-frame of the baryon.}  Dirac structures:
\begin{eqnarray}
V^{(3)}_{\rm conf}({\bf x_1}, {\bf x_2}, {\bf x_3})
&=& 
3\;a\;\;\frac{1}{4}\left[
\Id\tens\Id\tens\Id 
+
\gamma^0\tens\gamma^0\tens\Id
+ \textrm{cycl. perm.}
\right]\\
&&+\;\;
b\sum_{i<j} |{\bf x}_i-{\bf x}_j|\;\;
\frac{1}{2}\left[
-\Id\tens\Id\tens\Id 
+
\gamma^0\tens\gamma^0\tens\Id
+ \textrm{cycl. perm.}
\right].\nonumber
\end{eqnarray}
Here, the offset $a$ and the slope $b$ enter as free parameters in our
model.\\ In order to describe spin-dependent splittings, such as that
of the ground-state octet and decuplet, we adopt the explicitly
flavor-dependent 2-quark interaction\footnote{This interaction is
usually used in connection with the solution of the
$U_A(1)$-problem. In the framework of calculations for mesons \cite{kol00}
we have indeed shown that it yields the correct splitting of the
lowest meson nonet and in general of all low-lying meson states.
Moreover, we found \cite{loering01b} that the alternative QCD-based residual interaction --
the one-gluon exchange -- can be discarded on phenomenological
grounds.}, derived by 't~Hooft from instanton effects:
\begin{equation}
\label{tHooftPot}
V^{(2)}_{\rm 't~Hooft}({\bf x})
=
\frac{-4}{\lambda^3\pi^\frac{3}{2}}\;e^{-\frac{|{\bf x}|^2}{\lambda^2}}
 \left[\Id\tens\Id+\gamma^5\tens\gamma^5\right]{\cal P}^{\cal D}_{S_{12}=0}
\tens 
\left(
g_{nn}{\cal P}^{\cal F}_{\cal A}(nn)
+
g_{ns}{\cal P}^{\cal F}_{\cal A}(ns)
\right).
\end{equation}
Here, the four-fermion contact interaction has been smeared out by a
Gaussian function with effective range $\lambda$; the operator ${\cal
P}^{\cal D}_{S_{12}=0}$ in Dirac space projects on anti-symmetric
spin-singlet states, ${\cal P}^{\cal F}_{\cal A}(nn)$ and ${\cal
P}^{\cal F}_{\cal A}(ns)$ denote the projectors on
flavor-antisymmetric non-strange and non-strange--strange quark pairs.
Although the constituent quark masses $m_n$, $m_s$ as well as the
't~Hooft couplings $g_{\rm nn}$, $g_{\rm ns}$ and the range $\lambda$
can be related to standard QCD-parameters by instanton theory, we
determine these parameters by a fit to the experimental baryon
spectrum; for the values of the seven free parameters and a
consistency check of the QCD-relations we refer to
Refs. \cite{loering01b, loering01c}.\\
 
\begin{figure}[!t]
\epsfig{file={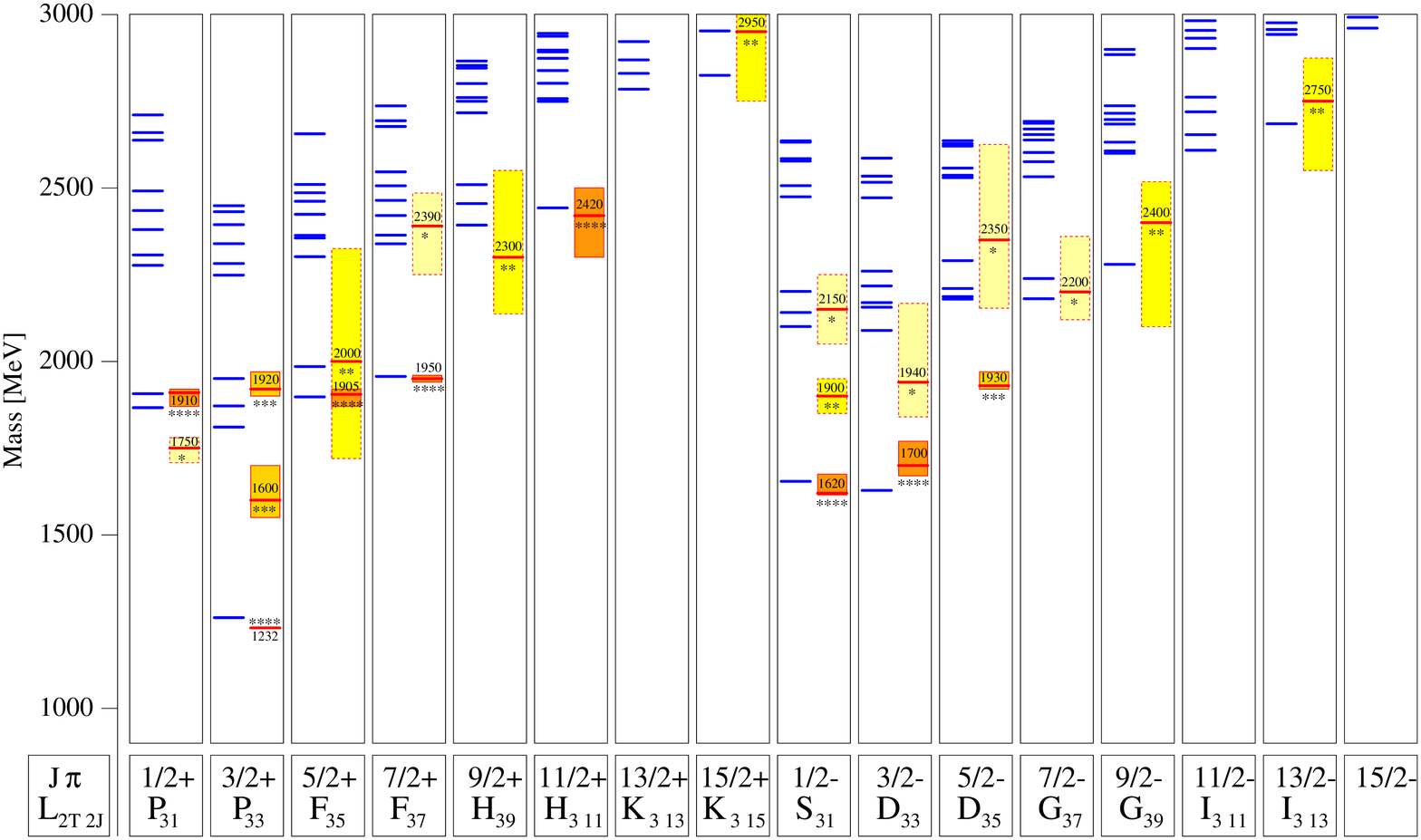},width=140mm}
\hfill
\parbox{.75\textwidth}{\epsfig{file={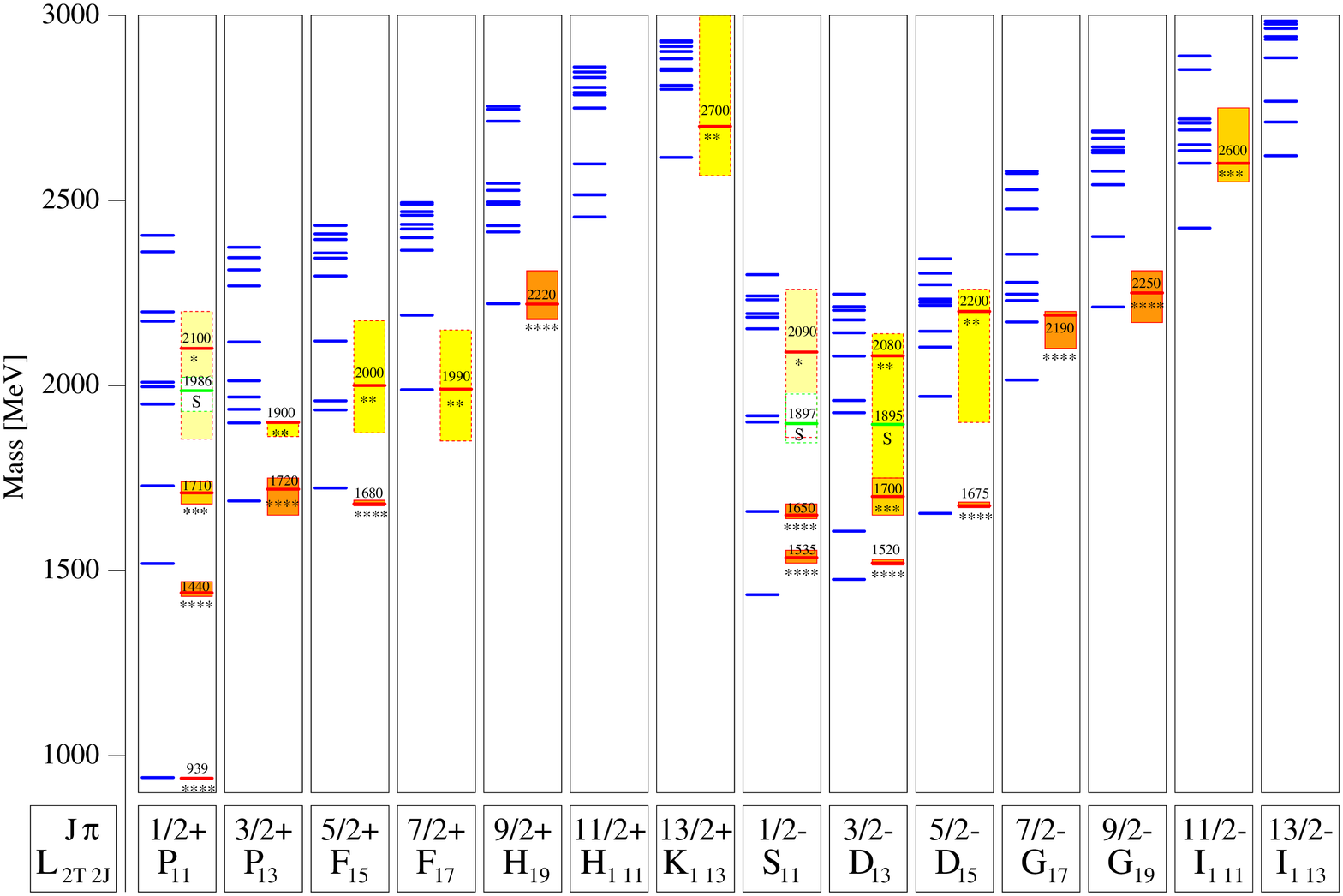},width=124mm}}
\hfill
\parbox{.24\textwidth}{
\caption{\small The complete $\Delta$- (above) and nucleon spectrum
(below): Our calculation in the left part of each column is compared
to the experimental spectrum \cite{PDG00} in the right part of each
column. Resonances are classified by total spin and parity
$J^\pi$. The experimental position is indicated by a bar, the
corresponding uncertainty by the shaded box which is darker for better
established resonances; the status of each resonance is additionally
indicated by stars.}
\label{fig:DeltaM2}}
\vspace*{-5mm}
\end{figure}

As 't~Hooft's interaction affects flavor antisymmetric qq-pairs only
this interaction does not act on flavor symmetric states, such as the
$\Delta$-resonances, which are thus determined by the dynamics of the
confinement potential alone.  Accordingly, the constituent quark
masses and the confinement parameters were determined by a fit to the
spectrum in this sector without any influence of the residual
interaction. In this manner a reasonable description of the complete
$\Delta$-spectrum including the linear Regge trajectory ($M^2\sim J$)
up to highest orbital excitations $J=\frac{15}{2}$ can be obtained,
see the upper part of Fig.~\ref{fig:DeltaM2}. 
However, the suspicious low position of the three negative-parity
``$3\hbar\omega$'' states quoted by the PDG \cite{PDG00} around 1900
MeV as well as that of the $\Delta\frac{3}{2}^+(1600)$ is not
accounted for (as in several other quark models). Since in our opinion
the experimental evidence of these negative parity states from
$\pi$--$N$-scattering is rather weak, a confirmation by complementary
(electro-/photo-production) experiments, such as the CLAS experiment
at CEBAF (JLab) or the Crystal Barrel experiment at ELSA (Bonn), seems
of great importance. 

The instanton-induced interaction does act on particular flavor octet
(and singlet) states.  The 't~Hooft coupling $g_{\rm nn}$ is adjusted
to the ground-state $N$-$\Delta$--splitting (and correspondingly
$g_{\rm ns}$ to the $\Xi^*$-$\Xi$-- and
$\Sigma^*$-$\Sigma$-$\Lambda$--splittings in the strange sector); then
all other states of the nucleon- (and the $\Lambda$-, $\Sigma$-,
$\Xi$-) spectrum are real parameter-free predictions. The lower part
of Fig. \ref{fig:DeltaM2} shows as an example our predictions for the
complete $N$-spectrum. As can be seen from this figure, our
predictions can nicely reproduce several features at least in
qualitative but mostly even in completely quantitative agreement with
the experimental findings.
\begin{figure}[!t]
\parbox{.3\textwidth}{
\caption{Instanton-induced effects in the nucleon spectrum. The curves
illustrate the change of the spectrum with increasing 't~Hooft
coupling $g_{nn}$ which finally is fixed to reproduce the correct
$N$-$\Delta$--splitting.}
\label{fig:nuclow}
}
\hfill
\parbox{.65\textwidth}{
 \epsfig{file={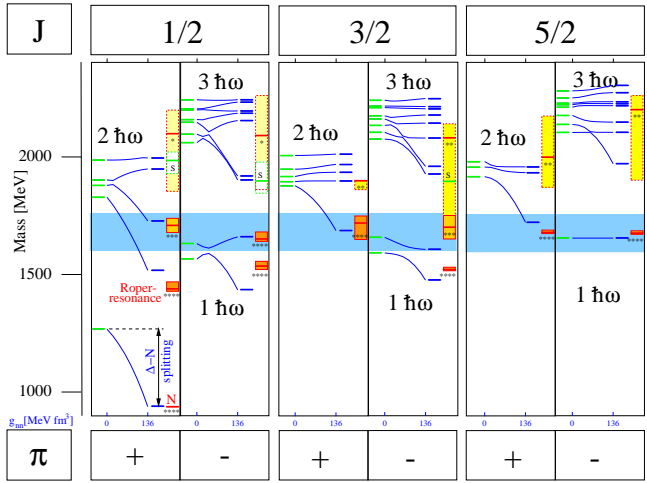},width=85mm}
}
\vspace*{-3mm}
\end{figure}

It is instructive to analyze the consequences of instanton effects for
the shape of the spectrum in some more detail.  Figure
\ref{fig:nuclow} illustrates the effect of 't~Hooft's force on the
lowest nucleon states. We find a selective lowering of some particular
states and once the coupling is fixed to reproduce the experimental
$N$--ground-state position we can automatically account very well for
the major spin-dependent mass splittings observed in the experimental
spectrum. For instance in the ``$2\hbar\omega$'' band exactly four
states are lowered relative to the other states such as required by
the experimental findings. In particular one finds that in this manner
the extremely low position of the prominent Roper resonance can be
accounted for quite naturally.  We should remark here that the
specific interplay of 't~Hooft's force, the relativistic effects and
the particular choice for the Dirac structure of the confinement
potential are essential to describe these effects quantitatively
\cite{loering01b}.

\begin{figure}[!t]
\parbox{.25\textwidth}{
\caption{\small Parity doublets for higher lying nucleon resonances.
See also caption to Fig.~\ref{fig:nuclow}\,.\label{fig:nuchigh}}
}
\hfill
\parbox{.75\textwidth}{
	\epsfig{file={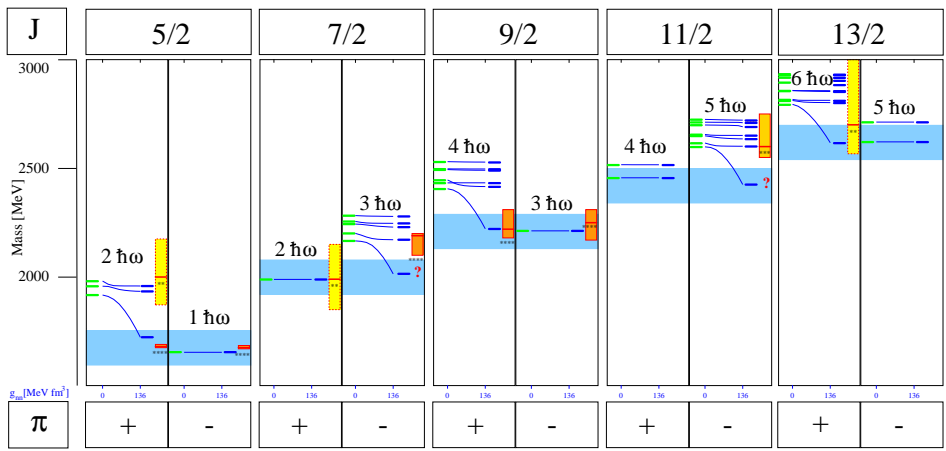},width=125mm}
}
\vspace*{-4mm}
\end{figure}

A very interesting feature of our relativistic quark model with
instanton induced forces is that it gives a natural dynamical
explanation of the so-called {\it parity doublets}: A glance at the
experimental $N$- and $\Lambda$-spectrum reveals a conspicuous
degeneracy of some states with the same spin and opposite parity.
Prominent, well-established, examples are
$N\frac{5}{2}^+\!(1680)$--$N\frac{5}{2}^-\!(1675)$,
$N\frac{9}{2}^+\!(2220)$--$N\frac{9}{2}^-\!(2250)$,
$\Lambda\frac{5}{2}^+\!(1820)$--$\Lambda\frac{5}{2}^-\!(1830)$\,.  In
the $\Sigma$--spectrum no clear indications of parity doublets is
found. In the literature these observations have been related to a
phase transition from the Nambu--Goldstone mode of chiral symmetry to
the Wigner--Weyl mode in the upper part of the baryon
spectrum\cite{Glozman:2000tk,Cohen:2001gb,Kirchbach:2001hj}\,.  In the
present model also this feature can be understood as an
instanton-induced effect: In general 't Hooft's force lowers those
sub-states of a major oscillator shell\footnote{Note that despite of
the linear confinement adopted here, the assignment of states to
oscillator shells still provides an adequate classification of states
with confinement alone.}  which contain so-called scalar diquarks,
i.e.  quark pairs with trivial spin and angular momentum. This is
found in particular for the high spin states in a given  $N\hbar\omega$ oscillator
shell, see Fig. \ref{fig:nuchigh}.
For given $N\hbar\omega$ the maximum total angular momentum for a
state containing such a scalar diquark is $J = (L_{\mbox{\footnotesize
max}}=N) + \frac{1}{2}$\,. 't Hooft's force lowers this state enough
to become almost degenerate with the unaffected spin-quartet state of
the adjacent oscillator shell with $N-1$, which has opposite parity
but the same total angular momentum: $J = (L_{\mbox{\footnotesize
max}}=N-1) + \frac{3}{2}$. In this way patterns of approximate parity
doublets for {\it all} lowest excitations in the sectors
$J=\frac{5}{2}$ to $J=\frac{13}{2}$ are formed {\it systematically}.
In the $N\frac{5}{2}^\pm$ and $N\frac{9}{2}^\pm$ sectors this scenario
is nicely confirmed experimentally by the well-established parity
doublets mentioned above:
$N\frac{5}{2}^+(1680)$--$N\frac{5}{2}^-(1675)$ and
$N\frac{9}{2}^+(2220)$--$N\frac{9}{2}^-(2250)$. In the $N\frac{7}{2}$
sector, however, the present experimental findings seem to deviate
from such a parity doubling structure due to the rather high position
of the $N\frac{7}{2}^-(2190)$.  Although this state is given a
four-star rating\cite{PDG00} an investigation of this sector with the
new experimental facilities would be highly desirable.  The same
mechanism explains approximate parity doublet structures also for
states with lower angular momentum as {\it e.g.}  the $N^*$ doublets
in the second resonance region around $\sim1700$ MeV with spins
$J^\pi=\frac{1}{2}^\pm$, $\frac{3}{2}^\pm$, and $\frac{5}{2}^\pm$ (see
Fig.~\ref{fig:nuclow}).  Observable consequences of this parity
doubling scenario should manifest in a different shape of
electromagnetic $\gamma^*p\rightarrow N^*$ transition form factors of
both members of a doublet due to their significantly different
internal structures: That member of the doublet, which is affected by
't Hooft's force (e.g. $N\frac{5}{2}^+$), exhibits a rather strong
scalar diquark correlation and thus its structure should be more
compact compared to its unaffected doublet partner
(e.g. $N\frac{5}{2}^-$) whose structure is expected to be rather
soft. Consequently, the transition form factor to the latter resonance
decreases faster than that to its doublet partner with the scalar
diquark contribution (see Fig.~3 of Ref. \cite{LoeMet01} for an
example). Finally, we should mention that in the strange sector (see
Ref. \cite{loering01c}) 't Hooft's force accounts in a similar way for
the prominent doublets of the $\Lambda$-spectrum. At the same time
instanton-induced effects are found to be significantly weaker in the
$\Sigma$-spectrum, thus explaining the fact that no clear experimental
indications of parity doublets are observed in this sector.\\

The investigations presented here have been restricted to the spectra
of baryons alone, which poses of course only a first (successful) test
of our model. Strong and electroweak decays as well as formfactors
pose additional stringent tests. In the present covariant framework
formfactors were already computed in the Mandelstam
formalism\cite{kretzsch01}; indeed we found that we can reliably
calculate such observables up to high momentum transfers.  Work on the
perturbative calculation of strong decays is more difficult but in
progress; first results are very promising, e.g. a {\it parameter-free} (!)
prediction \cite{merten01} of $\Delta\frac{3}{2}^+(1232)\rightarrow
N\pi$ yields a decay width $\Gamma = 109$ MeV in very good agreement 
with the experimental value $\Gamma=119 \pm 5$ MeV.
\acknowledgments{We would like to acknowledge the contributions of K.~Kretzschmar, D.~Merten
and H.-R.~Petry. We thank the Deutsche Forschungsgemeinschaft (DFG) for financial support.}


\end{document}